\documentclass[preprint,preprintnumbers,amsmath,amssymb]{revtex4}

\usepackage{epsfig}
\usepackage{hyperref}

\begin{document}
\title{Study on the Distribution Amplitude of the Scalar Meson $K_0^*(1430)$}
\author{Chen Wang$^{1,}$\footnote{Email: 371294924@qq.com}, Yuanyuan Ma$^1$\footnote{Email: 2243400179@qq.com }, Zhijun Wang$^1$\footnote{Email: 2587267208@qq.com
}and  Yanjun Sun$^1$ \footnote{Email: sunyanjun@nwnu.edu.cn}}
\address{$^1$ College of Physics and Electronic Engineering, Northwest Normal University}

\date{\today}

\begin{abstract}
Based on sum rules, we explore the twist-2 distribution amplitude of the  $K_0^*(1430)$ meson, treating it as the ground state of a quark-antiquark system. We posit that the spacetime distance $x$ should be infinitesimally close to the quark separation $ z$. By incorporating quark distance corrections, with $x^2 \approx z^2 \approx x z$, the calculated moments yield additional insights. Moreover, we employ light-cone sum rules to compute the form factors for the semi-leptonic decay process $B_s \rightarrow K$. The reliability of the computed distribution amplitude is confirmed through its comparison with the form factor.   \\


\end{abstract}

\maketitle

\section{introduction}
The $K_0^*(1430)$ meson is a scalar meson that serves as an intriguing subject of study due to its non-trivial structure. $K_0^*(1430)$ meson is usually viewed as the lowest energy state composed of a quark and an antiquark \cite{Chen:2021oul}.This description is well-supported in the literature, and the results of its mass calculated using the QCD sum rules method are quite consistent with experimental data\cite{Chen:2021oul}.

Another view is that the $K_0^*(1430)$ meson is a four-quark bound state composed of two quarks and two antiquarks\cite{Cheng:2019tgh}. This model explains why scalar mesons with masses below 1 \ GeV may have more complex structures. The four-quark state model can better explain certain experimental phenomena in some cases but is more complex theoretically, leading to less general acceptance.

Some researchers also take the view that the $K_0^*(1430)$ meson is a molecular state composed of two mesons\cite{Han:2013zg}. This model explains certain anomalous behaviors in the meson mass spectrum, but it receives less support when describing $K_0^*(1430)$, mainly due to its lower consistency with experimental data. Additionally, some studies propose that the $K_0^*(1430)$ meson might be a glueball state or a quark-gluon hybrid state\cite{Bharucha:2015bzk}. But so far, they receive only limited experimental support. 

The light-cone distribution amplitude (LCDA) of meson provides vital information about the momentum fractions carried by the quark and antiquark constituents, containing the non-perturbative aspects of QCD. Accurate knowledge of meson distribution amplitudes is crucial for both theoretical predictions and experimental validations, as they are fundamental to calculating hard scattering processes, exclusive decays, and form factors.  Therefore, computing meson distribution amplitudes with high precision is a central task in hadronic physics.

Several approaches have been developed to calculate the distribution amplitudes of mesons. One approach involves solving the Dyson-Schwinger Equations (DSEs)\cite{Xu:2018eii,Burden:1994yt,Maris:1997tm,Bashir:2012fs} within the framework of QCD. This method privides particular insights to the non-perturbative region and has been used to model the Pion and Kaon DAs. However, DSEs require sophisticated numerical techniques and approximations to handle the infinite tower of coupled integral equations, which may introduce uncertainties. Another approach is Lattice QCD (LQCD)\cite{Alford:2000mm,Sufian:2016hwn} which is a first principle method by discretizing space-time on a lattice. In LQCD calculations, quark and antiquark creation operators are used to study meson states. Smaller lattice spacings and larger lattice sizes can improve computational accuracy, but they also increase computational complexity\cite{Prelovsek:2013ela}. For heavy quarks, non-relativistic effects are minor, simplifying calculations, whereas for light quarks, relativistic effects cannot be ignored.

The DA can also be studied in light-front quark model\cite{Choi:2020xsr} through reasonable trial wave functions and spin-orbit coupling, combined with the variational principle and light-front quantization methods. This approach provides a direct caculation in the timelike region, avoiding the complexity of analytic continuation, and shows good agreement with experimental data\cite{Kang:2018jzg}. However, the computational complexity of this method is high, especially when dealing with high-energy regions and chiral limit conditions, as detailed theoretical analysis and extensive experimental data support is needed. The QCD sum rules\cite{Shifman:1978bx} method combines elements of both perturbative and non-perturbative QCD\cite{Khosravi:2022fzo}. Sum rules connect hadronic properties with QCD parameters by employing operator product expansion (OPE) and dispersion relations\cite{Ghahramany:2009zz}. This approach is less computationally intensive than Lattice QCD and more  fundamental than light-front holography\cite{Chen:2021oul}.

In this paper, we apply the sum rules method to  the distribution amplitude of the $K_0^*(1430)$ meson, considering the meson as a ground state of quark-antiquark pair. We  first derive the analytic form of the moment to obtain the distribution amplitude moments of  $K_0^*(1430)$ meson\cite{Yang:2005bv}. Subsequently, we utilize Gegenbauer polynomials to obtain the distribution amplitude and compare it with other results. Finally, we calculate the form factors for the $B_s\rightarrow K$ semileptonic decay process.  

\section{DISTRIBUTION AMPLITUDE OF $K_0^*(1430)$ MESON}

The Light-cone Distribution Amplitude (LCDA) of  $K_0^*(1430)$ meson  $\phi(x, \mu)$ can be defined as:
\begin{equation}\begin{gathered}
\left\langle K (P)\left|\bar{s}(x) \gamma_\mu  u(0)\right| 0\right\rangle= p_\mu \bar{f_K} \int_0^1 dxe^{iup\cdot x} \phi\left(u\right), \\
\left\langle K(P)\left|\bar{s}(0) u(0)\right|  0\right\rangle=  m _K\bar f_K,
\end{gathered}\end{equation}
here $ \bar{f}_{K}$  is the decay constant which refers to the decay amplitude of the meson in a certain decay mode,  $m_K$ is the mass of $K_0^*(1430)$ meson. Based on the conformal symmetry of the light-cone distribution amplitude, we can expand it in the form of Gegenbauer polynomials:
\begin{equation}
    \phi(u) = 6u(1-u)\left[1 + \sum_{n=1}^{\infty} B_n(\mu) C_n^{3/2}(2u-1)\right],
\end{equation}
where \(C_n^{3/2}(2x-1)\) are the Gegenbauer polynomials, and \(B_n(\mu)\) are the Gegenbauer coefficients that depend on the renormalization scale \(\mu\), reflecting the scale dependence dictated by the renormalization group equations. This expansion utilizes the fact that the Gegenbauer polynomials form a complete orthogonal basis under conformal transformations, allowing us to systematically incorporate the effects of QCD dynamics at different energy scales.  The Gegenbauer coefficients of the distribution amplitude can be represented by the moments of the DAs, since the moments can be defined as: 
\begin{equation}
  \left\langle\xi^n\right\rangle=\int_{\mathrm0}^1 d u(2 u-1)^n \phi(u) , 
\end{equation}
where $u=2x-1$. Substituting the distribution amplitude from Eq. (2) into the moments defined in Eq. (3), the relationship between the Gegenbauer coefficients and the moments is obtained as follows: 
\[
\begin{aligned}
& B_0=\xi^0\\
& B_1 = \frac{5}{3}  \xi^1, \\
& B_2 = \frac{7}{12} \left(\xi^0+5  \xi^2 \right), \\
& B_3 =- \frac{9}{4}  \xi^1+\frac{21}{4}  \xi^3 , \\
& B_4 = \frac{11}{24} \left(\xi^0 - 14  \xi^2+21  \xi^4 \right),\\
&...
\end{aligned}
\]
As a result, the key to studying the distribution amplitude lies in determining its moments.

Thus we construct the correlation function $\Pi(z, q)$ in order to derive the specific  $\xi$  moments : 
\begin{equation}\begin{aligned}
\Pi(z, q) & =i \int d^4 x e^{i q \cdot x}\left\langle\mathrm{0}\left|T\left\{J_n(x), J_{\mathrm{0}}(\mathrm{0})\right\}\right| \mathrm{0}\right\rangle \\
& =(z \cdot q)^{n+1} I\left(q^2\right)
\end{aligned}\end{equation}
with the interpolating currents
\begin{equation}\begin{aligned}
& J_n(x)=\bar{s}(x)(i \not z \cdot \overleftrightarrow{D})^n u(x) ,\\
& J_{\mathrm{0}}(\mathrm{0})=\bar{u}(\mathrm{0}) s(\mathrm{0}).
\end{aligned}\end{equation}
We now perform the operator expansion (OPE) for the correlator in the deep Euclidean region. The calculation is carried out in the framework of Background Field Theory (BFT) . By decomposing quark and gluon fields into classical background fields describing nonperturbative effects and quantum fields describingperturbative effects, BFT can provide clear physical images for the separation of long-range and short-range dynamics in OPE. So that the OPE of correlator (4) can be written as
\begin{equation}
    \begin{aligned}
\Pi_{2 ; K_{\mathrm{0}}^*}(z \cdot q) & =i \int d^4 x e^{i q \cdot x} \\
& \times\left\{-\operatorname{Tr}\left\langle\mathrm0\left|S_F^s(\mathrm0, x)(i \not z \cdot \overleftrightarrow{D})^n S_F^u(x, \mathrm0)\right| \mathrm0\right\rangle\right. \\
& +\operatorname{Tr}\left\langle\mathrm0\left|\bar{s}(x) s(\mathrm0)(i \not z \cdot \overleftrightarrow{D})^n S_F^u(x, \mathrm0)\right| \mathrm0\right\rangle \\
& +\operatorname{Tr}\left\langle\mathrm0\left|S_F^s(\mathrm0, x)(i\not z \cdot \overleftrightarrow{D})^n \bar{u}(\mathrm0) u(x)\right| \mathrm0\right\rangle \\
& +\cdots\}
\end{aligned}\end{equation}
where Tr indicates trace for the $\gamma$-matrix and color matrix, $S_F^s(\mathrm0, x)$ indicate the $s$-quark propagator from $x$ to 0, $S_F^u(x, 0)$ stands for the $u$-quark propagator from o to $x,(i z \cdot \overleftrightarrow{D})^n$ are the vertex operators from current $J_n(x)$. The specific forms of these matrix elements can be found in reference\cite{Colangelo:2000dp}. By substituting those corresponding formula into Eq. (6), the OPE of correlator (4), $I\left(q^2\right)$, can be obtained.

Then we will connect the OPE side with the moments, which can be given by expanding both sides of Eq. (1) 
\begin{equation}\begin{aligned}
\left\langle 0\left|\bar{s}(0)\not z(i z \cdot \overleftrightarrow{D})^n u(0)\right|  K(P)\right\rangle= & \bar f_K\left\langle\xi^n\right\rangle(z \cdot q)^{n+1},
\end{aligned}\end{equation}
where $D_\mu=\partial_\mu-i g_s T^A \mathcal{A}_\mu^A(x)(A=1, \ldots, 8)$ is the fundamental representation of the gauge covariant derivative.
By inserting a complete set of hadronic states into correlator (3) in physical region, whose hadronic representation can be read as

\begin{equation}
Im I^{had}\left(s\right)=\pi m_K \delta\left(s-m_K^2\right) \bar f_K{ }^2\left\langle\xi^n\right\rangle+Im I^{ope} \theta\left (s-s_0{ }\right) . 
\end{equation}
where $s_0$ is the threshold. The sum rules of moments reads:
\begin{widetext}\begin{equation}\begin{aligned}
\frac{1}{M^2} e^{-\frac{m_K^2}{M^2}} f^2 m_K\left\langle\xi^n\right\rangle 
& =\frac{3}{16 \pi^2} \frac{1}{(n+2)(n+1)}\left[3+(-1)^n+2 n\right] \int_0^{s_0} d s e^{-\frac{s}{M^2}}\left[(-1)^{n+1} m_u+m_s\right] \\
& +\left[\frac{m_s m_u}{2 M^2}+1+\frac{m_u^2(2 n+1)}{2 M^2}\right] \frac{\langle\bar{u} u\rangle}{M^2} \\
& +(-1)^{n+1}\left[-\frac{m_s m_u}{2 M^2}+1+\frac{m_s^2(2 n+1)}{2 M^2}\right] \frac{\langle\bar{s} s\rangle}{M^2}+\frac{2}{3} n g_s\langle\bar{s} G T \sigma s\rangle \frac{1}{M^4} \\
& +(-1)^{n+1} \frac{n}{3}g_s\langle\bar{u} G T \sigma u\rangle\frac{1}{M^4} .
\end{aligned}\end{equation}
\end{widetext}
In Eq. (9), $M$ is the Borel parameter, $\sigma_{\mu\nu}=\frac{i}{2}\left[\gamma_\mu\gamma_\nu\right]$, $m_u$ and $m_s$ are the current quark masses of the $u$ and $s$ quarks, $\langle\bar{u} u\rangle$ and $\langle\bar{s} s\rangle$ are the quark condensates, with $\langle\bar{s} s\rangle /\langle\bar{u} u\rangle=\kappa$. Here, $\kappa$ is a proportionality factor representing the ratio between the quark condensates of the $s$ quark and the $u$ quark. It quantifies the difference in vacuum condensate effects between different quarks, particularly when comparing lighter quarks like $u$ with heavier quarks like $s$. The quark-gluon mixed condensates are given by $ g_s\left\langle \bar{u} \sigma T G u\right\rangle$ and $ g_s\left\langle\bar{s} \sigma T G s\right\rangle$.

Above all we derive a reasonable analytic form for the moments of the DA. However, we believe its precision can still be improved. Traditional calculations often assume $z^2=0$, implying that the square of the quark separation distance is negligible compared to the square of the spacetime distance $x$. In which we find this approximation will be a bit slight, as in the case of short-distance high-energy interactions, the spacetime distance $x$ should be infinitesimally close to the quark separation $z$. With this approximation, we opt for $x^2 \approx z^2 \approx x z$ to correct the original calculation by adjusting the quark separation. Finally, the new sum rules for $\left\langle\xi^n\right\rangle$ read as follows:
\begin{widetext}
    \begin{equation}\begin{aligned}
 \left\langle\xi^n\right\rangle=& \frac{{M^2} e^{\frac{m_K^2}{M^2}}}{\bar{f}_K m_K}[ -\frac{3}{8 \pi^2}\left[5+3(-1)^n+2 n\right] \frac{1}{(n+1)(n+2)}] \int_0^{s_0} d s {e^{-\frac{s}{M^2}}\left[(-1)^{n+1} m_s+m_u\right]} \\
& +\frac{1}{M^2} \frac{2 n+1}{2 n+2}\left[\langle\bar{u} u\rangle+(-1)^{n+1}\langle\bar{s} s\rangle\right]+\frac{1}{3 \pi^2}\left[-(1)^{n+1} m_s+m_u\right] g_s\langle G^2\rangle\frac{1}{M^6} \\
& +\frac{2 n+1}{4 M^4}\left[m_u^2\langle\bar{u} u\rangle+(1)^{n+1} m_s^2\langle\bar{s} s\rangle\right]+\frac{m_s m_u}{4 M^4}\left[\langle\bar{s} s\rangle+\left(-1)^{n+1}\langle\bar{u} u\rangle\right]\right. \\
& -\frac{10+n}{24} \frac{1}{M^4}\left[g_s\langle  \bar{u} T G \sigma u\rangle+(-1)^{n+1}g _s\langle \bar{s} T G \sigma s\rangle\right]\\
&+\frac{16(n+1)}{81}  \frac{\pi}{M^6} {\left[m_sg_s\langle \bar s s\rangle^2+(-1)^{n+1} m_u g_s\langle \bar{u} u\rangle^2\right]+12 m_s(-1)^{n+1} \frac{1}{M^6}}\\
&+\frac{1}{12} m_s^2 g_s\left\langle G^2\right\rangle \frac{1}{M^4}],
\end{aligned}\end{equation}
\end{widetext}
while $ g_s \left\langle\bar{u} u\right\rangle^2$ and $ g_s \left\langle\bar{s} s\right\rangle^2$ represent the four-quark condensates. Next, we will numerically calculate the moments of the distribution amplitude. In the calculation of the Operator Product Expansion (OPE), the $S U_f(3)$ symmetry-breaking effect is considered. Specifically, the full $s$ quark mass effect in the perturbative part is preserved, and the $s$ quark mass corrections proportional to $m_s$ for condensate terms are calculated, considering $m_s \sim 0.1 \mathrm{\,\ GeV}$. Meanwhile, $m_u^2 \sim \mathrm{0}$ is adopted due to its small magnitude. The following values are used: $\langle\bar{u} u\rangle=-2.417\times 10^{-2} \mathrm{\ GeV}^3$,  $\langle\bar{s} s\rangle=\kappa\langle\bar{u} u\rangle \mathrm{\ with\ } \kappa=0.74$, $
\left\langle g_s \bar{u} \sigma T G u\right\rangle=-1.934 \times 10^{-2} \mathrm{\ GeV}^5$,  $\left\langle g_s \bar{s} \sigma T G s\right\rangle=\kappa\left\langle g_s \bar{u} \sigma T G u\right\rangle$,  $\left\langle g_s \bar{u} u\right\rangle^2=2.082\times 10^{-3} \mathrm{\ GeV}^6$, $\left\langle g_s \bar{s} s\right\rangle^2=\kappa^2\left\langle g_s \bar{u} u\right\rangle^2.$

Additionally, the scale $\mu=M$ is adopted in the sum rules Eq. (9).
Finally, we will determine the threshold parameter $s_0$. Since the $K_0^*(1430)$ meson is considered as a ground state meson composed of a quark-antiquark pair, and we selected the $K_0^*(1950)$ meson as its first excited state,  the threshold parameter is taken as the squared mass of the excited state particles, $s_0=1.944^2  \mathrm{\ GeV^2}$.  To enhance the reliability of the moment results, we select an appropriate range of Borel parameters. Our objective is to maximize the contribution of the non-perturbative term (i.e., the first term in Eq. (9)) while minimizing the contributions from the excited state and continuum spectrum terms (specifically, the dimension-six term). Given that the dimension-six term consistently accounts far less than 0.05\%, we choose the Borel window $M^2$ with the range of $1.4 \sim 1.8 \mathrm{\ GeV}^2$ and use the middle value as the input parameter. 
\begin{table}
\begin{ruledtabular}
    \centering
    \begin{tabular}{ccccc}
         &  $\xi^1$&  $\xi^2$&  $\xi^3$& $\xi^4$\\ \hline
          This work&  -0.337&  -0.116&  -0.224& -0.105\\
         SR\cite{Cheng:2005nb}& - 0.35&  —— &   -0.23& —— \\
         LF\cite{Chen:2021oul}&  -0.078&  -0.010&   -0.034& —— \\
         SR(LCHO)\cite{Huang:2022xny}&  -0.261&  0.0065&  -0.177& 0.0052\\
    \end{tabular}
    \caption{Our results for the first four moments of $\mathrm{K}_0^*(1430)$'s distribution amplitude, compared to Sum Rules(SR)\cite{Cheng:2005nb}, Sum Rules with LCHO model(SR(LCHO))\cite{Huang:2022xny} and Light Front model(LF)\cite{Chen:2021oul} .}
    \label{tab:my_label}
\end{ruledtabular}
\end{table}

\begin{figure}
    \centering
    \includegraphics[width=1\linewidth]{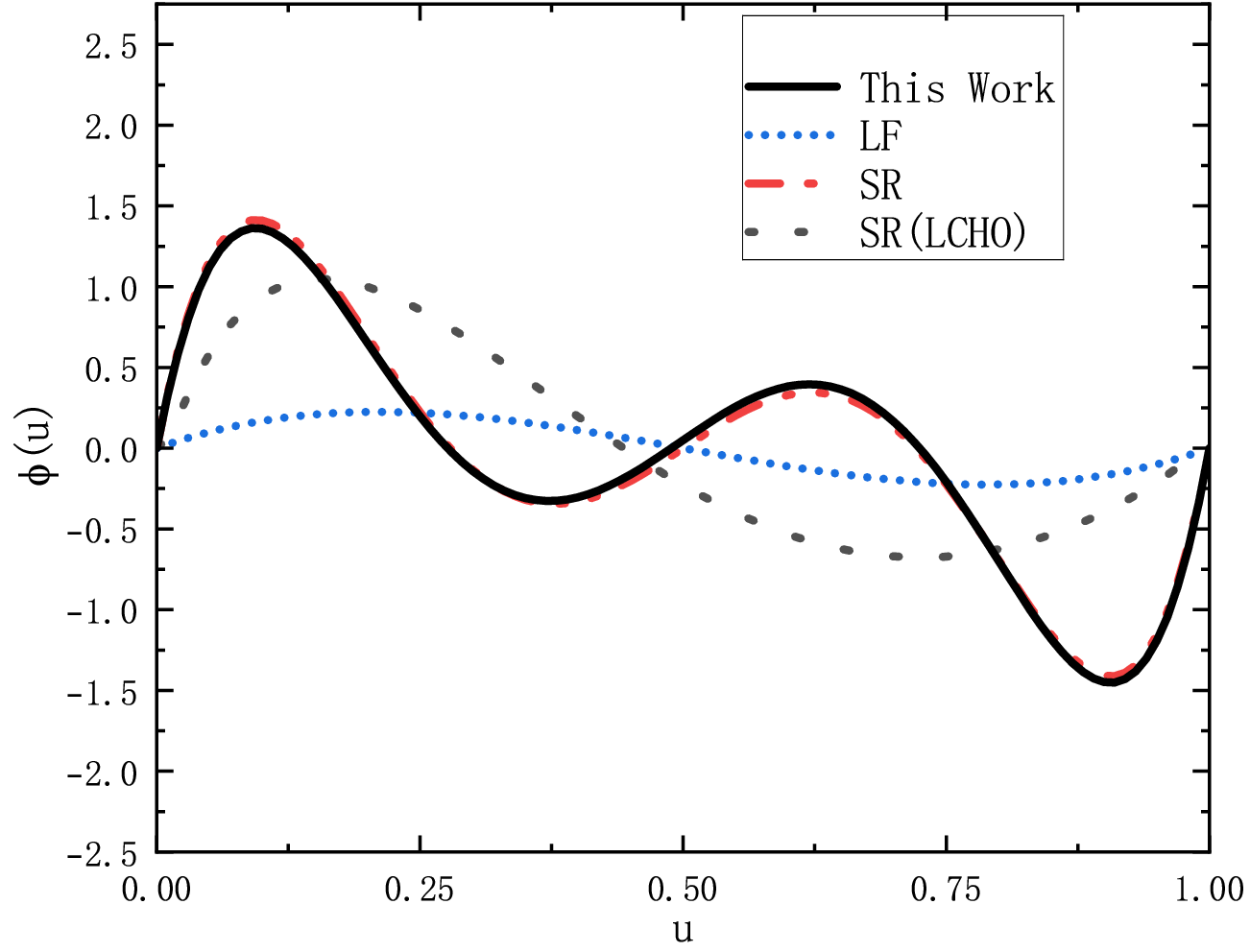}
    \caption{The distribution amplitude of $\mathrm{K}_0^*(1430)$ meson in this work, compared with Sum Rules(SR)\cite{Cheng:2005nb}, Light Front quark model(LF)\cite{Chen:2021oul}, Sum Rules with LCHO model(SR(LCHO))\cite{Huang:2022xny}. }
    \label{fig:enter-label}
\end{figure}
\begin{figure}
    \centering
    \includegraphics[width=1\linewidth]{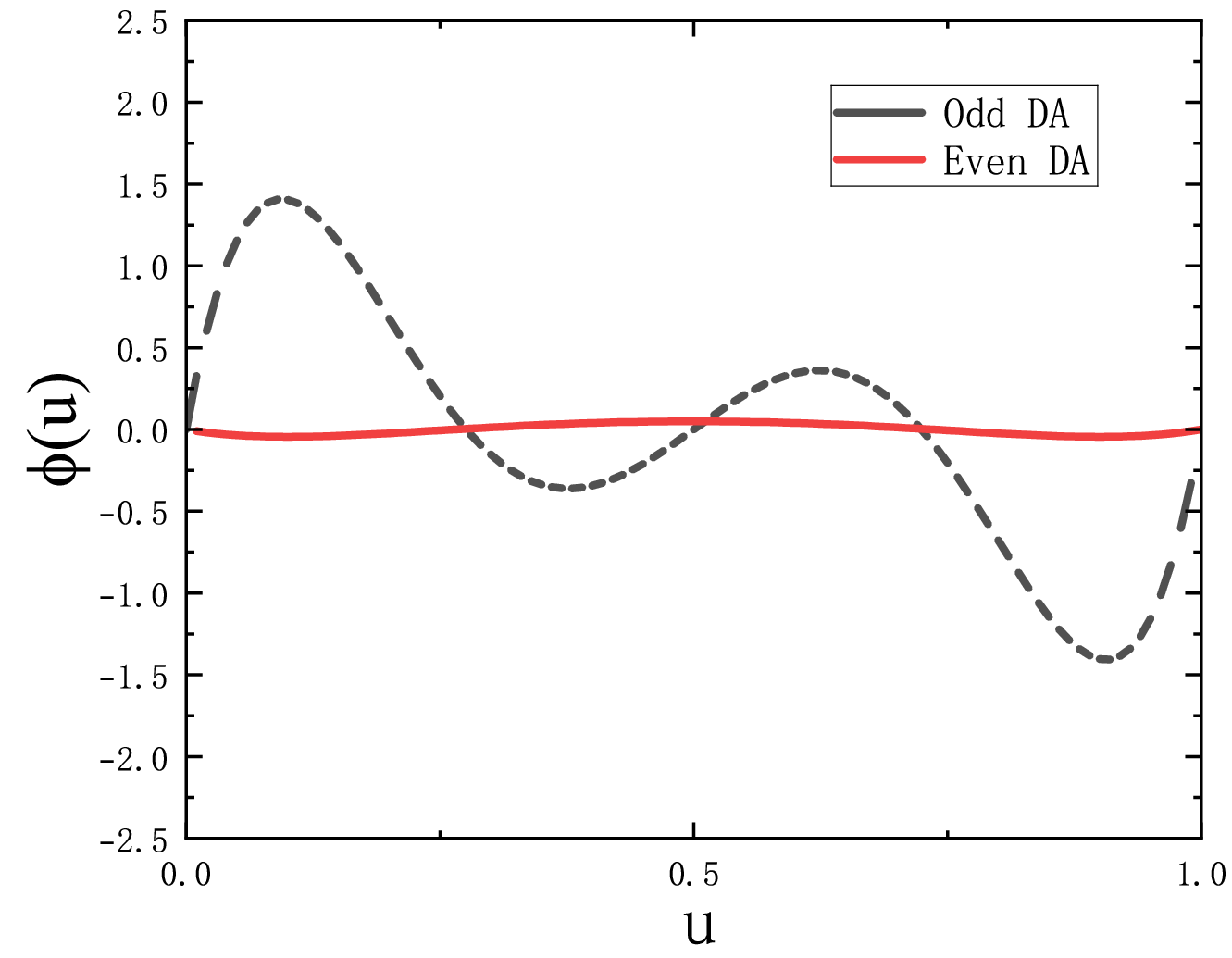}
    \caption{The odd-order and even-order distribution amplitudes of $\mathrm{K}_0^*(1430)$ meson.}
    \label{fig:enter-label}
\end{figure}
\begin{table}
\begin{ruledtabular}
    \centering
    \begin{tabular}{ccccc}
         &  $B_1$&  $B_2$&  $B_3$& $B_4$\\ \hline
          This work&  0.561&  0.024&  0.423& 0.019\\
         SR\cite{Cheng:2005nb}& - 0.435&  0.019&   -0.342& —— \\
         LF\cite{Chen:2021oul}&  -0.130&  -0.0130&   -0.005& —— \\
         SR(LCHO)\cite{Huang:2022xny}&  -0.57&  ——&  -0.42& ——\\
    \end{tabular}
    \caption{Our results for the first four Gegenbauer coefficients, compared to Sum Rules(SR)\cite{Cheng:2005nb}, Sum Rules with LCHO model(SR(LCHO))\cite{Huang:2022xny} and Light Front model(LF)\cite{Chen:2021oul} .}
    \label{tab:my_label}
\end{ruledtabular}
\end{table}
After calculating the moments, we use their values as input parameters to determine the Gegenbauer coefficients, which are then incorporated into the Gegenbauer polynomials to construct the distribution amplitude as is shown in FIG.1. From this figure, the distribution amplitude of the scalar meson $\mathrm{K}_0^*(1430)$ constructed using SR(LCHO) \cite{Huang:2022xny} and the light-cone distribution function\cite{Chen:2021oul}  method exhibits a bimodal pattern. In contrast, the distribution amplitude constructed using SR \cite{Cheng:2005nb} and the method proposed in this paper shows a four-peak pattern. Regarding the construction methods of the distribution amplitude, SR(LCHO)  employs the LCHO model, while the other three methods use the light-cone distribution function. The LCHO method, compared to the traditional light-cone distribution function, can eliminate the oscillation phenomenon when constructing the distribution amplitude with higher-order Gegenbauer coefficients. However, its normalization cannot be well guaranteed. The Gegenbauer coefficients calculated by the LF\cite{Chen:2021oul} method differ by orders of magnitude from those calculated by the sum rules, resulting in a much flatter distribution amplitude graph compared to the sum rule methods. Unlike the method in SR\cite{Cheng:2005nb}, both even and odd moments' contribution are contained in this paper. We separately constructed distribution amplitude plots based on odd-order and even-order Gegenbauer coefficients. It can be observed that the distribution amplitude plot constructed solely from the odd-order moments matches the shape found in the literature, but the contribution of the even-order moments leads to significant differences in the final plot shape(FIG.2).

\subsection{Transition Form Factors in the semileptonic decay process  $B_s\rightarrow K$}

\begin{figure}
    \centering
    \includegraphics[width=1\linewidth]{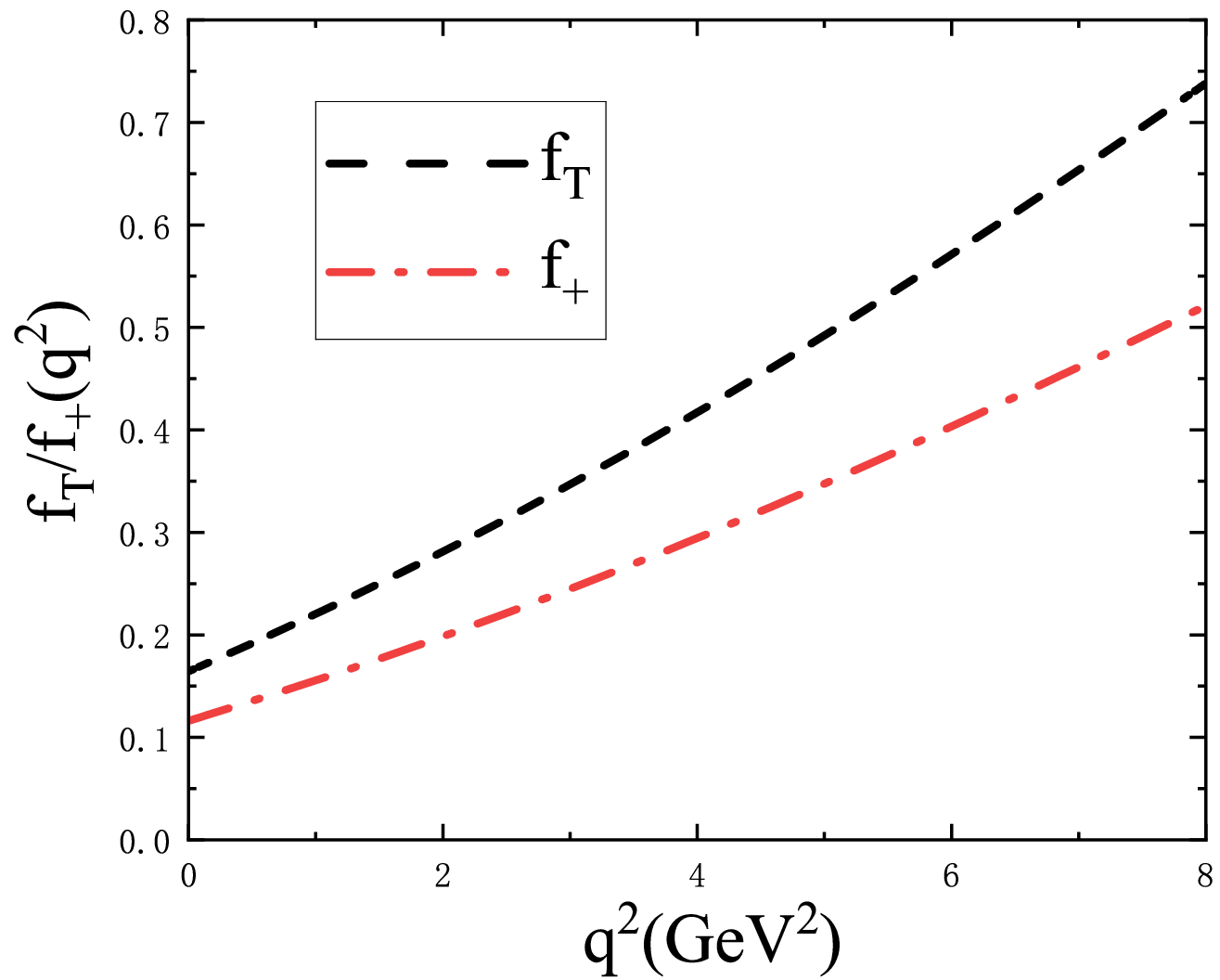}
    \caption{The dependence of the form factors for the $B_s\rightarrow K_0^*(1430)$ decay process on the transferred momentum $q^2$.}
    \label{fig:enter-label}
\end{figure}

Since the perturbative QCD method typically only calculates form factors at very small momentum transfers $q^2$, while lattice QCD shows good precision at large $q^2$, we choose  light-cone sum rules to calculate the form factors in the full range of $q^2$. Using the light-cone sum rule approach, the 
reference\cite{Sun:2010nv} provided the sum rules for the form factors in the semileptonic decay process $B_s \rightarrow K_0^*(1430)$:
\begin{equation}
\begin{aligned}
f_{+}^{B K}\left(q^2\right) & =-\frac{m_s+m_b}{m_B^2 f_B} m_b \int_{\Delta}^1 d u \frac{\phi(u)}{u} e^{FF}, \\
f_{-}^{B K}\left(q^2\right) & =\frac{m_s+m_b}{m_B^2 f_B} m_b \int_{\Delta}^1 d u \frac{\phi(u)}{u} e^{FF}, \\
f_T^{B K}\left(q^2\right) & =-\frac{m_s+m_b}{m_B^2 f_B}\left(m_B+m_K\right) \int_{\Delta}^1 d u \frac{\phi(u)}{u} e^{FF}.
\end{aligned}
\end{equation}
Here, $\phi(u)$ is the distribution amplitude of the $K_0^*(1430)$ meson, $m_s$ and $m_b$ are the mass of the $s$ and $b$ quark, $m_B$ is the mass of the $B_s$ meson, and $f_B$ is its decay constant.  The parameters $FF$ and $\Delta$ are defined as follows:
\begin{widetext}
    \begin{equation}
     \begin{aligned}
\Delta & =\frac{1}{2 m_K^2}\left[\sqrt{\left(s_0-m_K^2-q^2\right)^2+4\left(m_b^2-q^2\right) m_K^2}-\left(s_0-m_K^2-q^2\right)\right],\\
FF & =-\frac{1}{u M^2}\left[m_b^2+u(1-u) m_K^2-(1-u) q^2\right]+\frac{m_B^2}{M^2},
\end{aligned}\end{equation}
\end{widetext}
where $M^2$ is the Borel parameter. It is easy to see that the form factors obtained from the light-cone sum rules have simple relationships:
\begin{equation}\begin{aligned}
 f_{-}^{B K}\left(q^2\right)=&-f_{+}^{B K}\left(q^2\right), \\
 f_T^{B K}\left(q^2\right)=&\frac{\left(m_B+m_K\right)}{m_b} f_{+}^{B K}\left(q^2\right).
\end{aligned}\end{equation}
Therefore, by discussing $f_{+}^{B K}\left(q^2\right)$, the other two form factors can be easily obtained. 

To numerically calculate the form factor, we first determine the input parameters. The scale of the distribution amplitude will be input as $\mu=2.4\mathrm{\ GeV}$. The mass of $b$ quark, B meson and decay constant are selected as $m_b=4.8\mathrm{\ GeV}$, $m_B=5.368\mathrm{\ GeV}$, $f_B=0.23\mathrm{\ GeV}$. The threshold and the Borel parameter is chosed as $s_0^B=34\mathrm{\ GeV}^2$ and $M^2=14\mathrm{\ GeV}^2$ which is similar with reference\cite{Sun:2010nv}, resulting in the form factor dependencies on the transferred momentum, as shown in Fig.3. We compared our results with those obtained by other methods, as shown in FIG.3. It can be seen that our results are consistent with those in references\cite{Huang:2022xny,Sun:2010nv,Wang:2014vra,Khosravi:2022fzo,Yang:2005bv,Li:2008tk}, further verifying the reliability of the distribution amplitude calculated in the previous section.

\begin{table}\begin{ruledtabular}
    \centering
    \begin{tabular}{cccccll} 
    
Method&  This Work &  LCSR(1) &  SR & pQCD & LCSR(2) & LCSR(3)\\ \hline
         $f_+$ &  +0.115&  +0.10 &  +0.24 & -0.32 & +0.39 & +0.28\\ 
         $f_-$ &  -0.115&  -0.10 &  —— & —— & -0.24 & -0.10\\ 
         $f_T$ &  +0.164 &  —— &  —— & -0.41 &+0.43 &+0.32\\
    \end{tabular}
    \caption{Comparison of the form factors $f_+$, $f_-$ and $f_T$ at transferred momentum $q^2=0$ obtained using the light-cone sum rule (LCSR) in this work with results from light-cone sum rules (LCSR) \cite{Huang:2022xny,Sun:2010nv,Khosravi:2022fzo}, sum rules (SR) \cite{Yang:2005bv}, and perturbative QCD methods (pQCD) \cite{Li:2008tk}.}
    \label{tab:my_label1}
    \end{ruledtabular}
\end{table}

Regarding the results of the form factors at $q^2=0$, our results are consistent with those in Table.III. The discrepancies are mainly due to the different input distribution amplitudes, as the choice of threshold and the Borel window can lead to differences in the distribution amplitude results. Therefore, our calculation of the distribution amplitudes for the $K_0^*(1430)$ meson in twist-2 can be considered as a reliable result. The differences compared to the results in cite are due to the different currents selected in constructing the correlation function; the literature used an axial vector current, while we selected a chiral current, leading to differences in the positive and negative frequency form factor results. In contrast, cite used the same correlation function construction, but inserted the $K_0^*$ meson rather than the $B_s$ meson.

\section{Conclusion}

This paper investigates the twist-2 distribution amplitude of the scalar meson $K_0^*(1430)$ using the sum rules method. The $K_0^*(1430)$ is treated as a ground state of the quark-antiquark system, and we posit that the spacetime separation $x$ should be infinitesimally close to the quark separation $z$. By incorporating corrections to the quark separation distance, using the approximation  $x^2 \approx z^2 \approx x z$, we successfully compute the moments of the distribution amplitude, gaining additional insights into the meson's internal structure. Moreover, we employ light-cone sum rules to calculate the form factor for the semi-leptonic decay process $B_s\rightarrow K$. 
 
 For future perspectives, the results of this study provides a way to improve the precision of distribution amplitude calculations, which can be extended to other mesons. We anticipate obtaining more precise distribution amplitudes for various mesons through further research in the future.

\end{document}